\documentstyle[12pt,epsfig]{article}
\begin{document}
\begin{center}
{\large\bf  EXCESS SPECIFIC HEAT OF PTFE AND PCTFE AT LOW
TEMPERATURES: APPROXIMATION DETAILS}
\end{center}

\begin{center}
{ Nina B. Bogdanova$\, ^{a)}$, B.M. Terziyska$\,^{b)}$}
\end{center}

\noindent
{$\,^{a)}$Inst.of Nuclear Research,BAS, 72 Tzarigradsko choussee,
1784 Sofia, Bulgaria, {\it Email:nibogd@inrne.bas.bg}\\
$\, ^{b)}$Inst.of Solid State Physics,BAS, 72 Tzarigradsko choussee,
1784 Sofia, Bulgaria, {\it Email:terzyska@issp.bas.bg}
\noindent
\begin {abstract}
Approximation of the previously estimated excess specific heat $C^{excess}/T^{5}$ of
two
fluoropolymers, $PTFE$ and $PCTFE$,
is presented using Orthonormal Polynomial Expansion Method (OPEM).
 The new type of
weighting functions in OPEM involves the experimental errors in every point of
the studied thermal characteristic.
 The investigated temperature dependence of the function $C^{excess}/T^{5}$
 is described in the whole temperature ranges $0.4\div8 K$ and $2.5\div7
 K$ respectively for PTFE and PCTFE as well as in
 two subintervals $(0.4\div2) K$, $(2.5\div8) K$ for PTFE.
Numerical results of the deviations between the
evaluated $C^{excess}/T^{5}$ data
and their approximating values
are given. The  usual polynomial coefficients obtained by
orthonormal ones in our OPEM approach and the calculated in every point  absolute,
relative and specific sensitivities of the studied thermal characteristic are
proposed too.
The approximation parameters of this type thermal characteristic are
shown in Figures and Tables.
\end{abstract}

\bigskip
\noindent {\bf I.INTRODUCTION}
\medskip

The unusual thermal properties concerning the low-temperature
specific features of the heat capacities of the
polytetrafluoroethylene (PTFE) and polyclorothrifluoroethylene
(PCTFE) were considered in an earlier paper \cite {1}. The estimated there
excess specific heat over Debye contribution below 10 K of these fluoroplasts
was discussed in the frame of the recently developed
Soft-Potential Model (SPM).\\
The present study is devoted to the mathematical description of
the low-temperature excess specific heat of these semi-crystalline
polymers applying Orthonormal Polynomial Expansion Method
(OPEM)~\cite {2}.

\medskip
\noindent
{\bf II.EXCESS SPECIFIC HEAT DATA}
\medskip

The previous work \cite {1} clarifies several points concerning
the specific heat peculiarities observed in $C_{p}/T^{3}$
temperature dependencies of PTFE and PCTFE, a maximum appearing
around $T_{max}$= 4~K for both polymers and a found shallow
minimum for PTFE centered at $T_{min}=1~K$. Following the
Soft-Potential Model (SPM)~\cite {3,4,5,6}, supposing a
coexistence of acoustic phonons with quasi-localized low-frequency
(soft) modes in glasses and successfully applied by us to some
chalcogenide glasses \cite {7,8}, the specific heat data, taken
from the available calorimetric measurements at low-temperatures
 ($T~< 10$~K) for the studied fluoroplasts \cite{9,10}, were
 described in a paper  \cite{1}.
The $C_{p}$ components are:\\
i) $C_{p}^{TLS}$ - a linear contribution, described by double-well potentials,
conditioned by the thermal exitations of the tunneling state
(TLS); for PTFE it was established
to predominate at $T~\leq 0.2~K$. \\
 ii) $(C_{p}^{D})^{acou}$ - a cubic Debye contribution, evaluated and discussed in
details in a paper~\cite {1}. \\
 iii) $C_{p}^{exc(SM)}$ - an excess specific heat (a soft mode) contribution of the
quasi-harmonic exitations, described by single-well potentials. \\
$C_{p}^{exc(SM)}$ component was evaluated \cite {1} by difference between the measured
specific heat $C_{p}$ \cite {9,10} and the sum ($C_{p}^{TLS}$ +
$(C_{p}^{D})^{acou}$) as follows.
\begin{equation}
C_{p}^{exc(SM)}= C_{p}- [C_{p}^{TLS}+(C_{p}^{D})^{acou}], \label{(1)}
\end{equation}
The temperature dependencies of the three specific heat components in Eq.~\ref{(1)}
are:
\begin{equation}
C_{p}^{TLS} = C_{TLS}T, \hspace{0.3cm}
C_{p}^{D} = (C_{D})^{acou}T^{3}, \hspace{0.3cm}
C_{p}^{exc}= C^{exc(SM)}T^5. \label{(2)}
\end{equation}
The abbreviations of Two Level State, Debye and
excess specific heat (Soft Modes) are marked with $TLS$, $D$, and $exc(SM)$,
respectively.
Here,
$C_{TLS}$ was determined by the $C_{p}$ experimental data of Nittke et al~\cite {9};
$(C_{D})^{acou},$ the true elastic coefficient, was calculated~\cite {1} by the
macroscopic
parameters of the investigated materials and the average sound velocity $v_{s}$,
evaluated~\cite {1} for PTFE~(PCTFE) by available measurements of transversely and
longitudinally polarized 10 (5)~MHz ordinary sound waves between 4.2 and
140~(180)~K. Note that
the average sound velocity $v_{s}$ for both polymers changes its value about 0.3~\%
within
0 and 10~K. This fact allows us to accept up to 10~K a constant Debye coefficient of
the
Debye
specific
heat contribution by acoustic measurements .\\
The quantities $C_{D}$ up to 10~K and $C_{TLS}$ are independent of
the temperature. In accordance with the SPM, the softening of the
lattice vibrations leads to increasing of the density of states of
the harmonic oscillators with rising of their energy~$E$. This
increase appears to be proportional to the energy as $E^{4}$
leading to proportional to $T^{5}$ term in $C_{p}(T)$ at enough
low-temperatures, i.e. the quantity $C^{exc(SM)}$ is temperature
independent only in a narrow low-temperature range,
$0<T<1.5T_{min}$. It is worth to mention that the quantity
$C^{exc(SM)}$ evaluated for some glasses~\cite{11,12,13} confirms
this prediction of the SPM. For the studied polymers~\cite {1}
$T_{min}$ was estimated to be very close to 1~K. But in wider
low-temperature range the temperature dependencies of the
quantities $C_{p}^{exc(SM)}/T^{5}$ and the excess specific heat
$C_{p}^{exc(SM)}$ presented, respectively in Fig.1 and in the
inset of the {Fig.1}, are temperature dependent. As it can be
seen in the inset of Fig.1, the Soft Modes (SM) delocalized at T
= 8~K for PTFE and at T = 7~K for PCTFE.
\begin{figure}[htbp]
\centerline{\hbox{\epsfig{figure=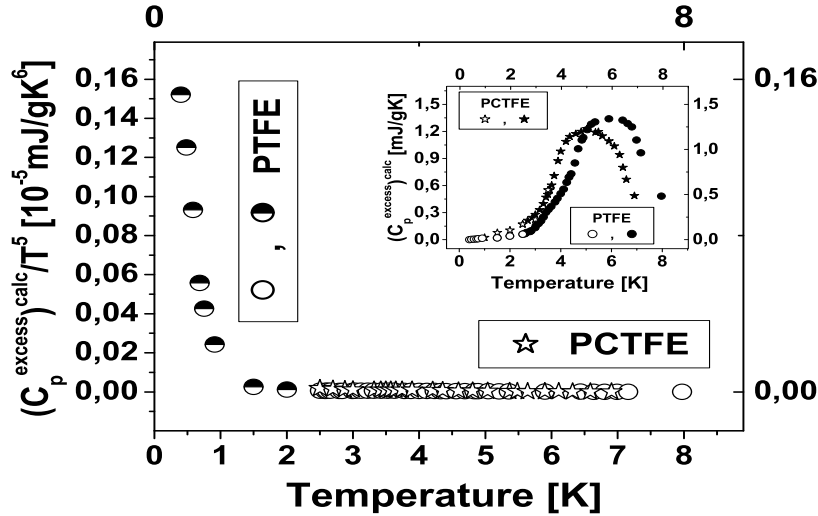,height= 8 cm,width= 12
cm}}}
\begin{center}
\caption{ Temperature dependencies of the excess specific heat
presented as $C^{exc}/T^{5}$; inset: Temperature dependencies of
the excess specific heat $C^{exc}$ $[ mJ/gK]$ of the PTFE and
PCTFE}.
\end{center}
\end{figure}
It is important to note that $C^{exc}/T^{5}$ function changes its
temperature behavior just at $T=(3/2)T_{min}$.

\medskip
 \noindent
 {\bf III. MATHEMATICAL APPROACH}
\medskip

 Our Orthonormal Polynomial Expansion
 Method(OPEM) and its applications
  in cryogenic thermometry are presented in papers
  \cite{2,14,15,16}.
   Some important features of OPEM concerning
    cryogenic thermometry at the approximation of
  thermometric characteristics of different type low-
  temperature sensors  are protected by a patent for an invention
  \cite{15}.
  Our OPEM is defined on Forsythe \cite{16}
 three-term relation for constructing orthogonal polynomials over
 discrete point set with arbitrary weights in the term of the least square method.
 The one-dimensional recurrence
 for  generation of orthonormal polynomials
 $\{\Psi_{k}^{(0)}$, $k=1,2,\ldots \}$ and their derivatives $\{\Psi_{k}^{(m)},
 m=0,1,2,\ldots \}$, in OPEM is:
 \begin{equation}
  \Psi_{k+1}^{(m)}(q )= \gamma _{k+1}[(q- \alpha_{k+1})\Psi_{k}^{(m)}(q)-\nonumber \\
 (1-\delta_{k0})\beta _{k}\Psi_{k-1}^{(m)}(q)+
   m\Psi_{k}^{(m-1)}(q)]
  \end{equation}
 OPEM is  a development of the Forsythe approach for receiving derivatives and
integrals
 with fourth term in the Eq. (3).
 The polynomials $\{\Psi_{k}^{(0)}\}$ satisfy
 the  orthogonality relations:
 $$\sum _{i=1}^{M}w_{i}\Psi_{k}^{(0)}(q_{i})\Psi_{l}^{(0)}
 (q_{i})=\delta _{kl}$$
 over the point set $\{q_{i}, i=1,2,\ldots M\}$ with weights
 $w_{i} =1/\sigma _{i}^{2} $ , depending on errors  $\sigma _{i}$ in every point.
 The approximating values $f^{appr}$ of the function and its $m$-th
 derivative $f^{(m)appr}$,$\{m=0,1,..\}$
 are calculated by
 \begin{equation}
 f^{(m)appr}(q)=\sum _{k=0}^{N}a_{k}\Psi_{k}^{(m)}(q)=\sum_{k=0}^Nc_{k}q^{k}.
 \label{(4)}
 \end{equation}
 The optimal degree $N$ of the approximating  polynomials  in Eq.(4)
 is selected by the algorithm, combining the following two criteria.
 First, the fitting curve  should lie in the error corridor of the dependent
 variable $(q_{j}, f_{j}^{exp}\pm{\sigma_{j}}, j=1,...M)$.
 \begin{equation}
 (f_{j}^{appr}-f_{j}^{exp})^{2}w_{j}\leq {1}.
 \end{equation}
 Second,  the
minimum $\chi^2$ should be reached.
\begin{equation}
\sum_{j=1}^M
{w_{j}(f_j^{appr}-f_{j}^{exp})^2}/(M-N-1)\rightarrow{min}.
\end{equation}
 When the first criterion is satisfied,
  the search of the minimum
  $\chi^2$ stops.
 The  development of the algorithm in the  biophysics with the total variance
formula for
 involving the errors in independent variable was
  published in a paper \cite{17}.
 The last version with  obtaining of usual
 $c_{k}$
 coefficients from orthogonal ones $a_{k}$  from Eq.~\ref{(4)} is developed
 in our work,
 RSI-2005~\cite{2}.

\bigskip
 \noindent
{\bf IV. APPROXIMATION RESULTS}\\
{\bf A. Orthonormal expansion}
\medskip

 The temperature dependence of the $C^{exc}/T^{5}$ function is described by
orthonormal polynomials
 in the whole temperature range $0.4\div8$~$[K]$ or in two subintervals
 $0.4\div2$~
 $[K]$,
  $2.5\div8$~$[K]$ for PTFE, and in the temperature range $2.5\div7$~ $[K]$ for
  PCTFE,
using the
 new type of weights, $W^{C^{exc}/T^{5}}$. The
studied thermal
 characteristic $C^{exc}/T^{5}$ and its sensitivities, absolute
$d(C^{exc}/T^{5})/dT$, relative
 $[1/(C^{exc}/T^{5})]d(C^{exc}/T^{5})/dT$ and specific,
$[T/(C^{exc}/T^{5})]d(C^{exc}/T^{5})/dT$,
 evaluated in every point, are shown in linear-linear plot in Figs.2
 and 3 for
 PTFE and PCTFE, respectively. It is worth to note that the subintervals for PTFE
are chosen by
 the temperature behavior of the specific sensitivity
 of this polymer (see Fig.2). \\
\begin{figure}[htbp]
\centerline{\hbox{\epsfig{figure=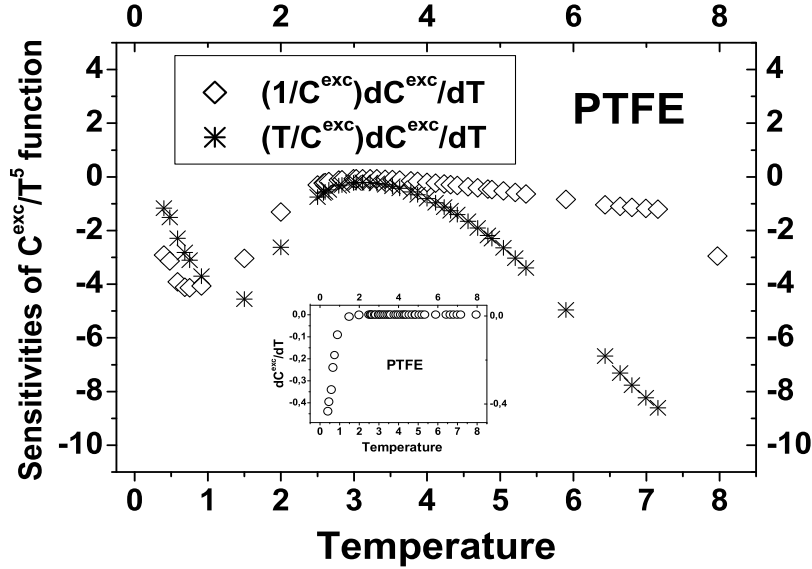,height= 9 cm,width=12
cm}}} \caption{Temperature dependencies of the relative
$[1/(C^{exc}/T^{5})]d(C^{exc}/T^{5})/dT$ $[K^{-1}]$ and specific
$[T/(C^{exc}/T^{5})]d(C^{exc}/T^{5})/dT$ $[-]$ sensitivities of
the PTFE; inset: Temperature dependencies of the absolute
sensitivity $d(C^{exc}/T^{5})/dT$.}
\end{figure}

\begin{figure}[htbp]
\centerline{\hbox{\epsfig{figure=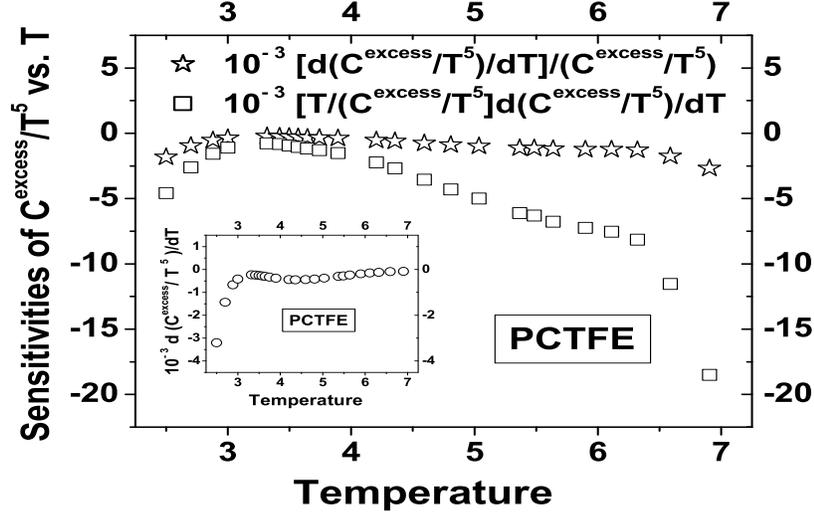,height= 8 cm,width=12
cm}}} \caption{ Temperature dependencies of the relative
$[1/(C^{exc}/T^{5})]d(C^{exc}/T^{5})/dT$ $[K^{-1}]$ and specific
$[T/(C^{exc}/T^{5})]d(C^{exc}/T^{5})/dT$ $[-]$ sensitivities of
the PCTFE; inset: Temperature dependencies of the absolute
sensitivity $d(C^{exc}/T^{5})/dT$ $[mJ/gK^{7}]$.}
\end{figure}
 By definition the weighting function $W^{C^{exc}/T^{5}}$ is $1/\sigma^{2}$, where
$\sigma^{2}$
 is a variance of the thermal characteristic $C^{exc}/T^{5}$ versus temperature $T$.
In our
investigation
 this variance is accepted to be, correspondingly square of the absolute heat capacity
 resolution $\Delta  C^{exc}_{acr}$, determined by the experimental specific heat
accuracy
 as follows:
 $(\Delta  C^{exc}_{acr})_{i}= 0.025 (C^{exc}/T^{5})_{i}$~~$[mJ/gK^{6}]$
 for the first approximating interval of PTFE and
 $(\Delta  C^{exc}_{acr})_{i}= 0.1 (C^{exc}/T^{5})_{i}$~~$[mJ/gK^{6}]$
 for both, the whole PCTFE approximating interval and the second approximating
interval of
 PTFE.
 Here the weights, $W^{C^{exc}/T^{5}}$  are expressed by the
relations:
\begin{equation}
 (W^{C^{exc}/T^{5}})_{i} = 1/(\Delta  C^{exc}/T^{5}_{acr})_{i}=
1600/(C^{exc}/T^{5})^{2}_{i}~~[mJ/gK^{6}]^{-2}
\end{equation}
for the first approximating interval of PTFE, or
\begin{equation}
 (W^{C^{exc}/T^{5}})_{i} = 1/(\Delta  C^{exc}_{acr})_{i}=
100/(C^{exc}/T^{5})^{2}_{i}~~[mJ/gK^{6}]^{-2}
\end{equation}
for the PCTFE and the second approximating interval of PTFE.
The deviations $\Delta  (C^{exc}/T^{5})_{i}$ between experimental and approximating
values of
 the excess specific heat are estimated in every point by the expression:
\begin{equation}
 \Delta  (C^{exc}/T^{5})_{i}= (C^{exc}/T^{5})_{i}^{exp} - (C^{exc}/T^{5})_{i}^{appr}
\end{equation}
 The temperature behavior of the calculated differences $\Delta  (C^{exc}/T^{5})_{i}$,
 the root mean square deviations $RMS^{C}$ and the mean absolute deviations
 $MAD^{C}$ respectively for PTFE and PCTFE excess specific heat
approximations
 are shown in the Figs.4,5 and 6.
\begin{figure}[htbp]
\centerline{\hbox{\epsfig{figure=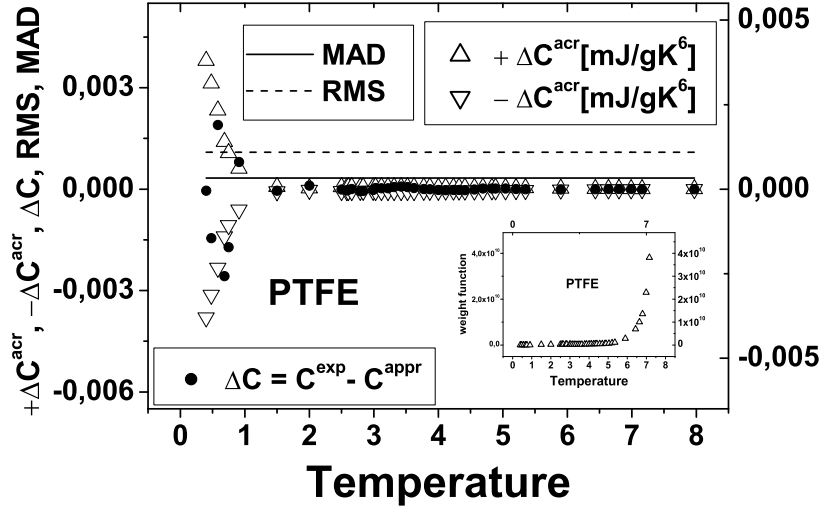,height= 8 cm,width=12
cm}}} \caption{Temperature dependencies of $\Delta
(C^{exc}/T^{5})$, $(\Delta (C^{exc}/T^{5})_{acr}$,
$MAD^{C^{exc}/T^{5}}$ and $RMS^{C^{exc}/T^{5}}$ of the PTFE for
the whole approximated temperature range;inset:Temperature
dependence of the weighting function $w(T)$ for FTPE .}
\end{figure}
\begin{figure}[htbp]
\centerline{\hbox{\epsfig{figure=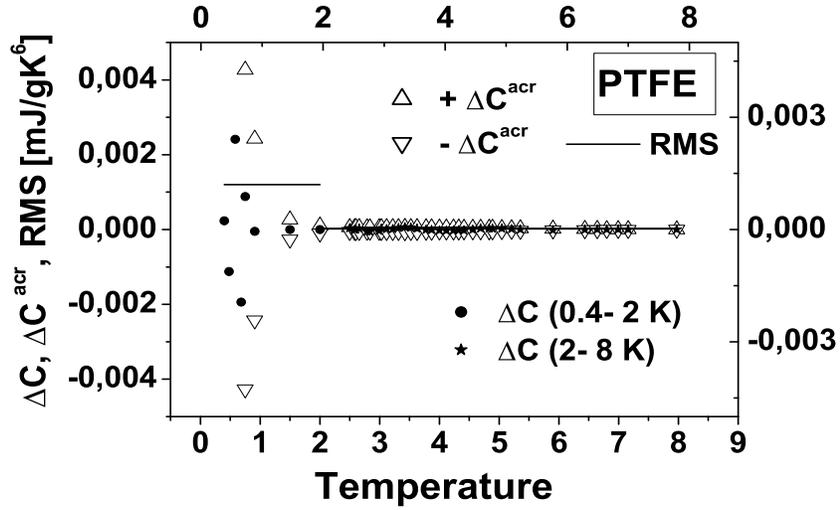,height= 8 cm,width=12
cm}}} \caption{Temperature dependencies of $\Delta
(C^{exc}/T^{5})$, $(\Delta (C^{exc}/T^{5})_{acr}$,
$MAD^{C^{exc}/T^{5}}$ and $RMS^{C^{exc}/T^{5}}$ of the PTFE for
two interval approximation.}
\end{figure}
\begin{figure}[htbp]
\centerline{\hbox{\epsfig{figure=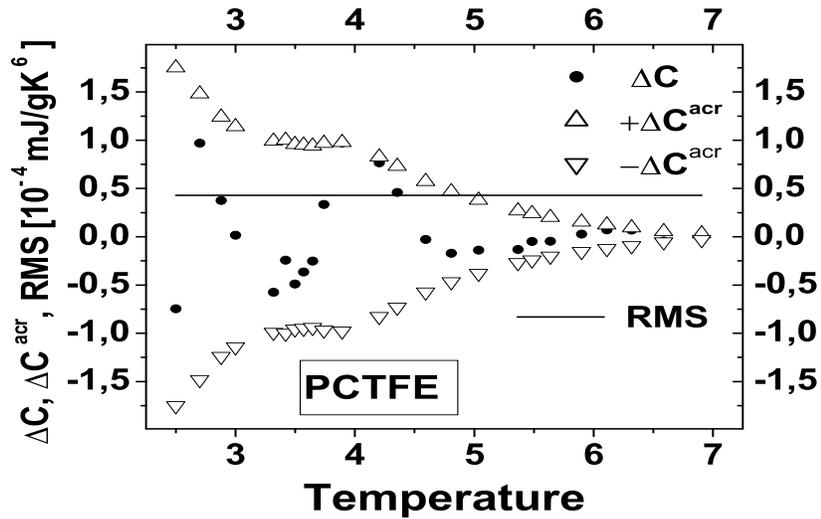,height= 8 cm,width=12
cm}}} \caption{Temperature dependencies of $\Delta
(C^{exc}/T^{5})$, $\Delta (C^{exc})_{acr}$, $MAD^{C^{exc}/T^{5}}$
and $RMS^{C^{exc}/T^{5}}$ of the PCTFE for the whole approximated
temperature range.}
\end{figure}
 The RMS deviation is
more
 popular. It is given in another our paper~\cite {2}. The characteristics MAD is
defined as follows:
\begin{equation}
MAD^{C}=\frac{1}{M}\sum_{i}\vert{\Delta (C^{exc}/T^{5})_{i}-
\overline{\Delta(C^{exc}/T^{5})}\vert},
 \end{equation}
 where $\overline{\Delta
(C^{exc}/T^{5})}=\frac{1}{M}\sum_{i}{\Delta(C^{exc}/T^{5})_{i}}$
 and $\Delta(C^{exc}/T^{5})_{i}$ are calculated  from Eq.(9).\\
 Following the cited criteria in Eqs.(5,6) the deviations $\Delta
(C^{exc}/T^{5})_{i}$
 are in the error corridor (see Eq.5). As a result of our approximation, the polynomial degree
for temperature range of PTFE  subintervals is lower than for the
whole temperature range. For $C^{exc}/T^{5}$ vs. $T$ approximation
the optimal
 degree $N$ as well as some main characteristics including the overall approximation
 characteristics: RMS and MAD of $\Delta C^{exc}/T^{5}$ vs. $T$ approximation,
the weighting
 function $W^{C}$ and a goodness of fit $\chi^{2}$ are presented in the
~{TABLE 1}.

\medskip
\noindent
{\bf B.Usual expansion obtained by orhonormal one}
\medskip

The next step in our approximation is done here.
 The  OPEM is extended by calculation of usual
 coefficients $\{{c}_{k}\}$ from orthonormal ones $\{{a}_{k}\}$
   using Eq.(4).
  This is carried out for the investigated intervals.
 The calculations are made for the $C^{exc}/T^{5}=f(T)$ descriptions, in two runs:
 first - in an interval [-1,1], and second - in the input intervals.
 Optimal values of usual polynomial degrees are chosen using two new criteria.
\renewcommand{\arraystretch}{0.8}
{\small
\begin{table}
\caption{ OPEM approximations of $C^{excess}/T^{5}$  for $PTFE$
and $PCTFE$}
\begin{tabular}{ccrccllc}
 \hline\hline
T&C&N&$W^{C}$&$RMS^{C}$&$MAD^{C}$&$\chi^{2}$\\
$[K]$&$[\mu J/gK^6]$&-&$[mJ/gK^6]$
&$[mJ/gK^6]$&~~~~-\\
 \hline\hline
$0.4\div2$&$152\div1$&6&$6.91\times{10}^{4}\div1.60\times{10}^{9}$&
$0.121\times10^{-2}$&$0.828\times10^{-3}$&1.38\\
$2.5\div8$&$1\div0.015$&4&$2.65\times 10^{8}\div4.44\times
10^{11}$&$0.265\times 10^{-4}$&$0.208\times
10^{-4}$&0.57\\
$0.4\div 8$&$152\div0.015$&8&$6.91\times 10^{4}\div4.44\times
10^{11}$&$0.109\times 10^{-2}$&$0.326\times10^{-3}$&0.95\\
\hline $2.5\div7$&$1.75\div0.031$&6&$3.26\times
10^{7}\div1.04\times 10^{11}$&
$0.428\times 10^{-4}$& $0.306\times 10^{-4}$&0.58\\
\hline\hline
\end{tabular}
\end{table}}

{\small
\begin{table}
\caption {Usual coefficients for $C^{excess}/T^{5}$~$[10^{-3}mJ/gK^{6}]$ vs. $T$
approximation}
\begin{center}
\begin{tabular}{|c||c|c|c||c|}
\hline\hline
\multicolumn{1} {|c||}{$material$}& \multicolumn{3}{c||}
{$PTFE$}&
\multicolumn{1}{c|}{$PCTFE$}\\
\hline \multicolumn{1}{|c||} {$Range T(K)$}& {$0.4\div2 $}&
{$2.5\div8 $}& {$0.4\div8 $}&
{$2.5\div7$}\\
\hline \multicolumn{1}{|c||} {$Range(C^{exc.}/T^{5})$}&
{$152\div1$}& {$1\div 0.015$}& {$152\div0.015$}&
{$1.75\div 0.031$}\\
\hline\hline
$c_{0}$& .022854    &  .003118   & .016067  & .182459 \\
$c_{1}$&-.001420    & -.009158   & -.063937 & -.667996 \\
$c_{2}$&  .105876   &   .011825  & .103755   & .997805   \\
$c_{3}$& -.582978   &   -.006389 & -.070355  &  -.769121  \\
$c_{4}$&  -.390617  &   .001190  &  -.008661  &  .320591  \\
$c_{5}$&  .845833   &     -      &  .048421   &   -.068068  \\
$ c_{6}$&  .696453  &     -      &  -.033867  &    .005687    \\
$ c_{7}$&    -      &    -       &  .010411   &    -         \\
$ c_{8}$&    -      &     -      &   -.001230  &     -        \\
\hline\hline
\end{tabular}
\end{center}
\end{table}}

The first criterion is:
 \begin{equation}
 \max|f_{i}^{exp}-f_{i}^{appr,u}|/f_{i}^{exp}\rightarrow{min},
  \end{equation}
 where $f^{appr,u}$ is the approximating function defined with
 usual expansion.
 The second criterion is:
\begin{equation}
   \Delta\{c_{k}\}_{1}^{N}\rightarrow{min},
   \end{equation}
where $\{\Delta c_{k}\}$ are inherited errors in usual
coefficients, defined in OPEM, discussed in a paper\cite{18}. In
the TABLE 2 the usual coefficients for $C^{exc}/T^{5}$ vs. $T$
approximation are presented.

In conclusion, the heat capacity components of the studied
fluoroplasts are well defined earlier \cite{1}  using the SPM. In
this study, the temperature dependencies of the excess specific
heat of PTFE and PCTFE below 10 K presented as
$C^{exc(SM)}/T^{5}$ are described by an Orthonormal Polynomial
Expansion Method (OPEM). An approximation of the above mentioned
thermal characteristic by usual polynomials, obtained by
orthonormal ones, is given too. The numerical results for
approximation characteristics as root mean square deviation, mean
absolute deviation and normalized $\chi$ square assure good
accuracy for calculated values of physical quantity in the input
temperature intervals. Our approach for description of this
characteristic proposes possibility for the future analyzing the
low-temperature excess specific heat of the materials with
glass-like behavior.

\end{document}